\documentclass[preprint]{sig-alternate}
\usepackage{graphicx}
\usepackage{stmaryrd}
\usepackage{mathbbol}
\usepackage{hyperref}
\usepackage[labelfont=bf,textfont=it]{caption}
\usepackage[all]{xy}

 \hypersetup{
   colorlinks,%
   citecolor=black,%
   filecolor=black,%
   linkcolor=black,%
   urlcolor=black}
\newcommand{\cou}[1]{\texttt{#1}}
\newcommand{\anyns}{\cou{any}}
\newcommand{\blns}{\cou{bl}}
\newcommand{\bnns}{\cou{bn}}
\newcommand{\qtyns}{\cou{qty}}

\newcommand{\pqns}{\cou{pq}}
\newcommand{\cvns}{\cou{cv}}
\newcommand{\cdns}{\cou{cd}}
\newcommand{\csns}{\cou{cs}}
\newcommand{\tsns}{\cou{ts}}
\newcommand{\iins}{\cou{ii}}
\newcommand{\ivltsns}{\cou{ivl$<$ts$>$}}
\newcommand{\ivlpqns}{\cou{ivl$<$pq$>$}}
\newcommand{\any}{\anyns\ }
\newcommand{\bl}{\blns\ }
\newcommand{\bn}{\bnns\ }
\newcommand{\qty}{\qtyns\ }

\newcommand{\pq}{\pqns\ }
\newcommand{\cv}{\cvns\ }
\newcommand{\cd}{\cdns\ }
\newcommand{\cs}{\csns\ }
\newcommand{\ts}{\tsns\ }
\newcommand{\ii}{\iins\ }
\newcommand{\ivlts}{\ivltsns\ }
\newcommand{\ivlpq}{\ivlpqns\ }

\newcommand{\evaluation}[1]{\ensuremath{\llbracket}{#1}\ensuremath{\rrbracket}}
\renewcommand{\epsilon}{\varepsilon}
\DeclareCaptionType[placement=b,within=none]{copyrightbox}
\toappear{Permission to make digital or hard copies of all or part of this work
  for personal or classroom use is granted without fee provided that copies are
  not made or distributed for profit or commercial advantage and that copies
  bear this notice and the full citation on the first page.  To copy otherwise,
  to republish, to post on servers or to redistribute to lists, requires prior
  specific permission and/or a fee. \\ Copyright \copyright 2010 MGRID B.V.
\href{http://www.mgrid.net}{\tt www.mgrid.net}}

\begin{document}
\title{Adding HL7 version 3 data types to PostgreSQL}
\numberofauthors{3}
\author{
\alignauthor
Yeb Havinga\\
	\affaddr{MGRID}\\
        \affaddr{Oostenburgervoorstr. 106-114}\\
        \affaddr{1018MR Amsterdam}\\
        \email{y.t.havinga@mgrid.net}
\alignauthor
Willem Dijkstra\\
	\affaddr{MGRID}\\
        \affaddr{Oostenburgervoorstr. 106-114}\\
        \affaddr{1018MR Amsterdam}\\
        \email{w.p.dijkstra@mgrid.net}
\alignauthor
Ander de Keijzer\\
	\affaddr{University of Twente}\\
        \affaddr{Institute of Technical Medicine}\\
        \affaddr{7500AE Enschede}\\
        \email{a.dekeijzer@utwente.nl}
}
\date{\today}

\maketitle

\begin{abstract}
The HL7 standard is widely used to exchange medical information
electronically. As a part of the standard, HL7 defines scalar communication
data types like physical quantity, point in time and concept descriptor but also
complex types such as interval types, collection types and probabilistic
types. Typical HL7 applications will store their communications in a database,
resulting in a translation from HL7 concepts and types into database
types. Since the data types were not designed to be implemented in a relational
database server, this transition is cumbersome and fraught with programmer
error. The purpose of this paper is two fold. First we analyze the HL7 version
3 data type definitions and define a number of conditions that must be met, for
the data type to be suitable for implementation in a relational database. As a
result of this analysis we describe a number of possible improvements in the
HL7 specification. Second we describe an implementation in the PostgreSQL
database server and show that the database server can effectively execute
scientific calculations with units of measure, supports a large number of
operations on time points and intervals, and can perform operations that are
akin to a medical terminology server. Experiments on synthetic data show that
the user defined types perform better than an implementation that uses only
standard data types from the database server.
\end{abstract}

\section{Introduction}
The HL7 version 3 standard specifies how to exchange medical information
electronically. The original intent of the HL7 standard is to provide a
framework for designing medical messaging applications, though there is an
increasing number of initiatives that apply HL7 methodology and structure to
their entire medical information system. This has been noticed by the HL7
organization and in 2008 the RIM\footnote{RIM is the HL7 version 3 {\it Reference
    Information Model}.} Based Application Architecture (RIMBAA) working group
has been founded\cite{rimbaa}.

Part of the HL7 specification is the data type specification. Since differences
in data type implementations between programming environments are common, HL7
makes no assumptions about available data types, but defines it's own
instead\cite{hl7dat}. The definitions of the individual types are structured in
an object oriented fashion; generic types are extended by more specific types,
with all types inheriting from a top abstract type. The top type has a notion
of a flavor of null; this allows HL7 developers to provide information about a
value when it is null. HL7 defined data types include scalar types like
integer, string, point in time, but also complex types and templated types such
as interval types, collection types and types expressing probabilities. The HL7
data types in its second revision called R2 has become an ISO standard in
2008\cite{iso21090}.

\subsection*{Example data type}
Data type {\tt Boolean} specializes top abstract type {\tt ANY}; it inherits
{\tt ANY}'s ability to express a nullflavor and extends that with the boolean
value domain. This is defined in the following fashion:

\begin{verbatim}
type Boolean alias BL specializes ANY
    values(true, false) {
        BL and(BL x);
        BL not;
        literal ST.SIMPLE;
        BL or(BL x);
        BL xor(BL x);
        BL implies(BL x);
};
\end{verbatim}

Assertions are then made on how the particulars of this data type should be
interpreted, such as the definition $x.not$ as negation of $x$\/:
\begin{verbatim}
invariant(BL x) {
    true.not.equal(false);
    false.not.equal(true);
    x.isNull.equal(x.not.isNull);
};
\end{verbatim}

The mapping of the HL7 data types can be done in a number of ways. PostgreSQL
pioneered the following approach: the definition of User-Defined Types that
uniquely encapsulate the HL7 data types, allowing full access to the types and
methods on those types from SQL\cite{designofpostgres}
\cite{Stonebraker86inclusionof} \cite{udtpagina34.11}.

\begin{figure}
\centering
\includegraphics[width=8.5cm]{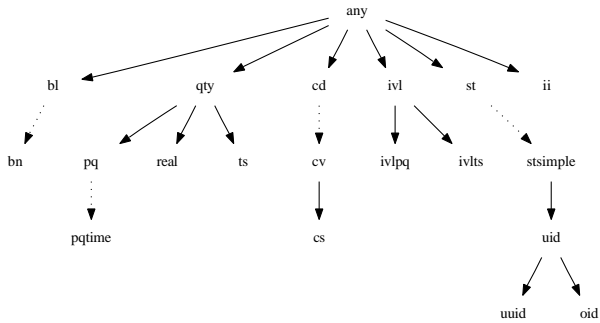}
\caption{Part of the HL7 data type hierarchy}
\label{fig:hl7typehier}
\end{figure}

HL7 data types were not designed to be implemented in a relational database, and
it is no surprise that some facets of the data types are a challenge to match to
relational database theory:

\begin{itemize}
\item{The hierarchy of types; this implies that table columns of a general type
  must be able to hold instances of its subtypes. See figure
  \ref{fig:hl7typehier} for a subset of the hierarchy.}
\item{Not all data types have a string literal format; making it hard to work
  out what the domain of those types are.}
\item{The top type in the hierarchy has the ability to express that it is null
  in a particular flavor. For instance: a data value can be null with flavor {\tt asked but
    unknown}.}
\end{itemize}

We investigate to what extent the HL7 data type definitions can be matched with
relational database theory, define a set of conditions for a data type that must
be satisfied before it can be implemented and implement a subset of the
data types in PostgreSQL.

\section{Preliminaries}

\begin{table*}
\centering
\caption{HL7 NullFlavors}\vskip.5em
\begin{tabular}{|l|l|l|l|}\hline
{\bf Level}&{\bf Symbol}&{\bf Meaning}&{\bf Description}\\ \hline\hline
1&{\bf ni}&no information&Default and most general exceptional value\\ \hline
\hskip5mm2&{\bf inv}&invalid&Value not permitted in constrained domain\\ \hline
\hskip10mm3&{\bf oth}&other&Actual value not permitted in constrained domain\\ \hline
\hskip15mm4&{\bf ninf}&negative infinity&Negative infinity on numbers\\ \hline
\hskip15mm4&{\bf pinf}&positive infinity&Positive infinity on numbers\\ \hline
\hskip10mm3&{\bf unc}&unencoded&Information not encoded (yet)\\ \hline
\hskip10mm3&{\bf der}&derived&Actual value must be derived\\ \hline
\hskip5mm2&{\bf unk}&unknown&Proper value is applicable but not known\\ \hline
\hskip10mm3&{\bf asku}&asked but unknown&Information sought but not found\\ \hline
\hskip15mm4&{\bf nav}&temporarily unavailable&Expected to be available later\\ \hline
\hskip10mm3&{\bf qs}&sufficient quantity&Quantity not known but non-zero\\ \hline
\hskip10mm3&{\bf nask}&not asked&Information not sought\\ \hline
\hskip10mm3&{\bf trc}&trace&Greater than zero but too small for quantification\\ \hline
\hskip5mm2&{\bf msk}&masked&Information available but not provided\\ \hline
\hskip5mm2&{\bf na}&not applicable&No value applicable in context\\ \hline
\end{tabular}
\label{nullflavors}
\end{table*}

In this section we review relevant portions of relational database theory, as
well as the parts of the specification of HL7 data types that are relevant to
this paper.

\subsection{HL7 Data type definitions}
\label{hl7definitions}
HL7 Data types are defined in a hierarchy, with the data type \any as most
general data type. Instances of data types are called \emph{data values}. The
hierarchy of types that are discussed in this paper is shown in figure
\ref{fig:hl7typehier}.

For every data type \emph{properties} are defined. A property is referred to by
its name. The \emph{domain} of a property is the set of values a property can
have. Properties may have arguments. For instance, the {\tt plus} property of
one integer argument requires another integer as argument.

\paragraph*{Any}
The data type \any is the most generic data type. All other data types extend \any 
and inherit functions defined on \anyns. \any is abstract, and is only used to
define a function interface and common properties.

The HL7 notion of NULLs, called \emph{nullflavors}, are defined on \anyns.
Nullflavors are specified to extend the domain of the data type, allowing
nullflavor symbols in all HL7 data type literals. The nullflavors are defined in
a hierarchy, which coincides in meaning with the ordering on degree of
knowledge described in \cite{fourvalued}. The most general nullflavor is no
information, or {\tt ni}, all other nullflavors add meaning to that. Table
\ref{nullflavors} shows all 15 and their hierarchy. The level code in the first
column encodes the position in the hierarchy; the nullflavor in each row is
subsumed by the closest nullflavor above it with a lower number. For instance,
trace / {\tt trc} is subsumed by unknown / {\tt unk} but not by invalid / {\tt
  inv}.

The \any data type has several properties that are true if the data value is of
a certain nullflavor, like \cou{nonnull}, \cou{isnull} and \cou{unknown}. Also
the properties \cou{equal} and \cou{identical} are defined to take another data
value as argument and have \cou{bl} as domain.

\begin{table*}
\centering
\caption{AND truth table}\vskip.5em
\begin{tabular}{|l|l|l|l|l|l|l|l|l|l|l|l|} \hline
{\bf AND}&{\bf asku}&{\bf false}&{\bf inv}&{\bf msk}&{\bf na}&{\bf nask}&{\bf
  nav}&{\bf ni}&{\bf oth}&{\bf true}&{\bf unk}\\ \hline\hline
{\bf asku}&{\sl asku}&false&ni&ni&ni&unk&asku&ni&ni&asku&unk\\ \hline
{\bf false}&false&{\sl false}&false&false&false&false&false&false&false&false&false\\ \hline
{\bf inv}&ni&false&{\sl inv}&ni&ni&ni&ni&ni&inv&inv&ni\\ \hline
{\bf msk}&ni&false&ni&{\sl msk}&ni&ni&ni&ni&ni&msk&ni\\ \hline
{\bf na}&ni&false&ni&ni&{\sl na}&ni&ni&ni&ni&na&ni\\ \hline
{\bf nask}&unk&false&ni&ni&ni&{\sl nask}&unk&ni&ni&nask&unk\\ \hline
{\bf nav}&asku&false&ni&ni&ni&unk&{\sl nav}&ni&ni&nav&unk\\ \hline
{\bf ni}&ni&false&ni&ni&ni&ni&ni&{\sl ni}&ni&ni&ni\\ \hline
{\bf oth}&ni&false&inv&ni&ni&ni&ni&ni&{\sl oth}&oth&ni\\ \hline
{\bf true}&asku&false&inv&msk&na&nask&nav&ni&oth&{\sl true}&unk\\ \hline
{\bf unk}&unk&false&ni&ni&ni&unk&unk&ni&ni&unk&{\sl unk}\\ \hline
\end{tabular}
\label{11valuedand}
\end{table*}

\begin{table*}
\centering
\caption{OR truth table}\vskip.5em
\begin{tabular}{|l|l|l|l|l|l|l|l|l|l|l|l|} \hline
{\bf OR}&{\bf asku}&{\bf false}&{\bf inv}&{\bf msk}&{\bf na}&{\bf nask}&{\bf
  nav}&{\bf ni}&{\bf oth}&{\bf true}&{\bf unk}\\ \hline\hline
{\bf asku}&{\sl asku}&asku&ni&ni&ni&unk&asku&ni&ni&true&unk\\ \hline
{\bf false}&asku&{\sl false}&inv&msk&na&nask&nav&ni&oth&true&unk\\ \hline
{\bf inv}&ni&inv&{\sl inv}&ni&ni&ni&ni&ni&inv&true&ni\\ \hline
{\bf msk}&ni&msk&ni&{\sl msk}&ni&ni&ni&ni&ni&true&ni\\ \hline
{\bf na}&ni&na&ni&ni&{\sl na}&ni&ni&ni&ni&true&ni\\ \hline
{\bf nask}&unk&nask&ni&ni&ni&{\sl nask}&unk&ni&ni&true&unk\\ \hline
{\bf nav}&asku&nav&ni&ni&ni&unk&{\sl nav}&ni&ni&true&unk\\ \hline
{\bf ni}&ni&ni&ni&ni&ni&ni&ni&{\sl ni}&ni&true&ni\\ \hline
{\bf oth}&ni&oth&inv&ni&ni&ni&ni&ni&{\sl oth}&true&ni\\ \hline
{\bf true}&true&true&true&true&true&true&true&true&true&{\sl true}&true\\ \hline
{\bf unk}&unk&unk&ni&ni&ni&unk&unk&ni&ni&true&{\sl unk}\\ \hline
\end{tabular}
\label{11valuedor}
\end{table*}

\paragraph*{Boolean (BL)}
The data type \bl has as domain the truth values \emph{true} and \emph{false}
with literal form {\tt true} and {\tt false}. It inherits from \anyns, which
means that (1) its literal form is extended with the nullflavors. The
nullflavors {\tt ninf}, {\tt pinf}, {\tt unc}, {\tt der}, {\tt qs} and {\tt
  trc} are not allowed on the boolean data type. (2) \bl inherits all properties
of \any such as the boolean null properties like \cou{isnull} from \anyns. (3)
It must also provide an implementation for the \cou{equal} property. The
equality of two operands is true when both operands are non-null and have the
same value. It adds the following properties {\tt not}, {\tt and}, {\tt or},
{\tt xor}, {\tt equal} and {\tt implies}, as expected with their usual
meaning. When one or two of the operands have a nullflavor, the operators
behave like Codds three valued semantics\cite{coddextending}. Since the number
of nullflavors allowed on \bl is 9, there are $11^2$ possible
combinations. According to the HL7 data type specification, when operands are
two different nullflavors, the result is the nullflavor `that is the first
common ancestor of the two nullflavors'. The ancestor relation is the usual,
except that in this context it is also reflexive; if both operands are the same
nullflavor, the result is the same nullflavor. This results in a 11 valued
logic as is shown in tables \ref{11valuedand} and \ref{11valuedor}. (4)
Finally, since \bl is a proper subtype of \anyns, table attributes of type \any
can hold instances of type \blns.

\paragraph*{Boolean NonNull (BN)}
The data type \bn is a constrained version of the \bl data type so that it is not
null. Constrained versions of types are called \emph{flavors} in the HL7
data type specification. In figure \ref{fig:hl7typehier} flavors can be
distinguished from normal specialized types by the dotted arrows.

\paragraph*{Quantity (QTY)}
The data type \qty specializes \any and like \anyns, it is an abstract type. No
data values can be just of type \qty without belonging to a proper subtype of
\qtyns. \qty subsumes ordered data types and hence introduces the properties
      {\tt lessorequal}, {\tt lessthan}, {\tt greaterthan}, {\tt
        greaterorequal}.  Also it adds the addition and subtraction operators
      {\tt plus} and {\tt minus} on operands of the same data type.

\paragraph*{Real Number (REAL)}
\label{real}
The real number is a scalar magnitude, which is used whenever quantities are
measured, estimated, or computed from other real numbers. The literal form is
decimal, where the number of significant decimal digits is known as the
precision.  Conforming implementations are not required to be able to represent
the full range of real numbers. The standard declares the representations of
the real value space as floating point, rational, scaled integer, or digit
string, and their various limitations to be out of the scope of the
specification.

\paragraph*{Physical Quantity (PQ)}
\label{pq}
One of the most practical HL7 types is \pqns, which is a complex data type
consisting of a unit and a value property, where the unit is a string constant
that conforms to the \emph{Unified Code for Units of Measure} or UCUM
specification \cite{ucum}, and the value is the quantity. Example units are
{\tt m} for meter, {\tt J} for Joule, {\tt mm[Hg]} for pressure. The UCUM
specification consists of a few hundreds units, and defines 7 base units which
are {\tt m} for meter, {\tt g} for gram, {\tt s} for second, {\tt rad} for
radian, {\tt K} for Kelvin, {\tt C} for Coulomb and {\tt cd} for candela.
Every other unit can be expressed in terms of these base units together with
metric prefixes like {\tt k} for kilo and {\tt m} for milli. For instance, {\tt
  mm[Hg]} can be expressed as $\tt m^{-1}.g.s^{-2}$. Example {\pqns}s are {\tt
  10 ml} and {\tt 0.5 kg/m2}.

The \emph{canonical value} property of a \pq $x$ is a \pq $y$ such that $x = y$
and the unit of $y$ consists only of base units. Two {\pqns}s \emph{compare}
if their canonical values have the same base units. For instance, {\tt ml} and
{\tt dm3} compare, but {\tt mm} (millimeter) and {\tt m3} (a cubic meter) do
not.

Two units are \emph{equal} if the values of their canonical values are
equal. Two units are \emph{identical} if both the units and values are the
same. For instance, {\tt 1 m} and {\tt 100 cm} are equal, but are not
identical.

\paragraph*{Point in time (TS)}
\ts is a specialization of \qty that defines a point on the axis of natural
time. Since there is no absolute zero-point on the time-axis a \ts is measured
in the amount of time that has elapsed since some arbitrary zero-point called
the \emph{epoch}. The current standard supports only the Gregorian calendar,
with \ts measurements in quantities comparable to any \pq with base unit {\tt
  s}. Every \ts has a precision that specifies the number of significant
digits in the calendar expression. The calendar expression can hold year,
month, day, hour, minute, second, fractional seconds and timezone.

\paragraph*{Interval (IVL$<$T$>$)}
The interval definition is a template that turns any quantity type into an
interval type of that quantity. Two specific intervals are discussed below.

\paragraph*{Interval of point in time (IVL$<$TS$>$)}
The \ivlts interval form has seven different literal forms, listed in table
\ref{intervals}. While few forms define a start and end of an interval, other
define only a single point in time or only a period of time without specific
starting point. Brackets in the interval, centerwidth and width forms show if
the interval is inclusive or exclusive of the boundary. An example: the
interval between the first of January, 13:12:51 and the 31st of January, time
15:56:29 in 2008 is {\tt [20080101131251;\-20080131155629]}. An entire calendar
month must be denoted using an open high boundary, like so: {\tt
  [20010101;\-20010201[}.

The promotion of \ts to \ivlts uses the precision of the source \ts to define
the width of the new \ivltsns. An example: the promotion of \ts {\tt 20010131}
yields an \ivlts of {\tt [20010131000000;20010201000000[}.

\begin{table}
\centering
\caption{Interval forms for TS}\vskip.5em
\begin{tabular}{|l|l|l|l|} \hline
{\bf Name}&{\bf Example}\\ \hline\hline
Interval form    & {\tt [20080101131251;20080131155629]} \\ \hline
Comparator form  & {\tt <20080101}                       \\ \hline
Centerwidth form & {\tt 20010115135108 [10s]}            \\ \hline
Width form       & {\tt [10d]}                           \\ \hline
Center form      & {\tt 20010101}                        \\ \hline
Any form         & {\tt ?200101?}                     \\ \hline
Hull form        & {\tt 20010101..20010131}              \\ \hline
\end{tabular}
\label{intervals}
\end{table}

\paragraph*{Interval of physical quantities (IVL$<$PQ$>$)}
\ivlpq also has 7 different literal forms. It lacks a hull form,
but it gains a dash form like so: {\tt 3ml - 5ml} which is equivalent to
{\tt[3ml;5ml]}.

\paragraph*{Concept Descriptor (CD)}
Controlled vocabularies play an important role in medical informatics. In the
HL7 data type specification, the concept descriptor data type provides
terminological functionality.

A \cd is a reference to a concept defined in a code system, such as LOINC, ICD
or SNOMED-CT. Properties of \cd include a code, an original text that served as
the basis of the coding, and zero or more translations of the concept into
multiple code systems. The code can be an atomic code, an elementary concept
directly defined by the reference code system, for instance {\tt C41.9} for the
concept `Bone and articular cartilage, unspecified' in ICD-10. A code can also
be an expression in some syntax defined by the referenced code system to
facilitate post-coordination. Post-coordination is the usage of a set of codes
to explain a single concept that does not have a code in the vocabulary
yet. SNOMED-CT is an example of a code system that supports post-coordination.
Many code systems have an explicit notion of concept specialization and
generalization. The {\tt implies} property on \cd takes another \cd as argument
and returns true iff the argument is a generalization of the data value.

\paragraph*{Coded Value (CV)}
A \cv is a flavor of \cd, such that there are no translations and only a single
concept is allowed. \cv is used when only a single code value is required.

\paragraph*{Coded Simple Value (CS)}
A \cs is a specialization of \cvns. It is coded data in its simplest form, where
everything but the code itself is predetermined by the context in which the \cs
data value occurs. Hence the literal form of a \cs consists of only the code
itself.

\paragraph*{Instance Identifier (II)}
An instance identifier uniquely i\-dent\-ifies a thing or object. Properties of \ii
include the root and the extension. The root is character string that is either
an DCE Universal Unique Identifier (UUID) or an ISO/IEC 8824:1990 Object
Identifier and is a unique identifier that guarantees the global uniqueness of
the \ii data value. The extension is a character string that is unique in the
namespace of the root. The extension may be null. The literal form of \ii is
defined to be the root, followed by a colon, followed by the extension.

\subsection{Database definitions}
The database definitions we need are the usual, with two exceptions. First, we
need the distinction between the set of constants and the object domain. Though
this distinction is rarely made in relational database literature, we need it in
the discussion of semantics of nullflavors, as well as in section
\ref{conditions} where a number of conditions are described that data types must
satisfy, in order to be suitable for implementation in a relational
database. Also, since this paper is concerned with data types only, we do not formally
define relation schemes, instances and attributes, but only a set of types.

{\bf dom} is a countably infinite set of individual objects. {\bf con} is a
countably infinite set of constant symbols. Let {\bf typ} denote the set of all
data types. $Dom$ is a mapping from {\bf typ} to {\bf dom}. $Con$ is a mapping
from {\bf typ} to {\bf con}. If $A$ is a data type, $Dom(A)$ is called the
\emph{domain of} $A$ and $Con(A)$ is called the \emph{literal form of} $A$.
Constants describe objects. The meaning \evaluation{a} $\subseteq$ {\bf dom} of
a constant $a$ is a mapping from {\bf con} to {\bf dom}, and is the set of all
objects described by $a$. $a$ is a better description than $a'$, if
\evaluation{a} $\subset$ \evaluation{a'}.
The \emph{unique names assumption} means that no two different constant symbols
denote the same object. 
\subsection{Partial information}
The concept of a null value in relational databases has been widely discussed
in database literature because of its problematic nature.
One of the problems is the interpretation of a null value; common
meanings found in database literature are \emph{unknown}, a value is applicable
and exists but is at present not known, and \emph{nonexistent}, a value does
not exist. Zaniolo\cite{zani} proposes a third interpretation called \emph{no
 information} and shows that a sound relational algebra can be constructed
with this interpretation, though meaning is lost due to the more general nature
of \emph{no information}.  In \cite{coddextending} a three valued semantics for
boolean operations with unknown values was given. \cite{fourvalued} describes
problems that occur in many valued semantics and introduces ordering partial
information with increasing degree of knowledge. This idea is expanded by
others, for a complete reference we refer to \cite{libkinpartial}.

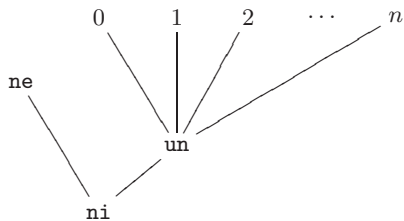
\begin{figure}
\centering
\begin{displaymath}
\xymatrix @ - 10pt {
      & 0 \ar@{-}[ddr]
        & 1 \ar@{-}[dd]
          & 2 \ar@{-}[ddl] & \ldots
            & n \ar@{-}[ddlll] \\
{\tt ne} \ar@{-}[ddr] & & & & &\\
& & {\tt un} \ar@{-}[dl] & & &\\
& {\tt ni} & & & &\\
}
\end{displaymath}
\caption{Degree of knowledge ordering}
\label{fig:degorder}
\end{figure}

What is the meaning of a null value with respect to domains?  \cite{zani}
extends each domain to include the distinguished symbol {\tt ni}. \cite{bjo91}
proposes an ordering on values of domains based on the number of real-world
objects a described by a symbol, such as a null value. For example, for the
domain of natural numbers and the null values {\tt ni}, {\tt un} and {\tt ne},
meaning no information, unknown and nonexistent, respectively, the ordering is
shown in figure \ref{fig:degorder}. This picture shows a number of
things. First of all it captures the idea that the \emph{unknown} null value
means that any of the natural numbers could be the actual value. In other
words, one of the naturals is equal to \emph{unknown} on this data type. This
does not hold for {\tt ne}, which does not match any natural number. {\tt ni}
is the most general null value. The actual value is not known and it is even
not known whether an actual value exists. {\tt ne} as well as every natural
number represent \emph{perfect knowledge}, in the sense that the information
can not be improved. These correspond to the the leafs of the tree.

\section{Analysis}
\penalty100
In this section we review the HL7 data type definitions from the standpoint of
relational database theory. Our goal is to match HL7 data type definitions to
relational database concepts. We discuss relevant portions of some of the
implemented constructs and data types.

The concept of data type \emph{properties} was described in section
\ref{hl7definitions}. Though the HL7 standard does not specialize properties,
we make the distinction between two different kinds:
\begin{enumerate}
\item{\emph{compositional properties} are the intrinsic components of a
  data type. For instance, \cou{value} is a compositional property of the
  data type \pqns.}
\item{\emph{relational properties} define relations on data values of certain
  types. For instance, the \cou{equal} relational property of the \pq data type
  takes as argument another \pq and thereby defines a relation on the
  powerdomain of \pq data values.}
\end{enumerate}
A data type property is either a compositional property or a relational
property, but never both. This distinction is needed in the next section.

\subsection{Conditions}
\label{conditions}
Let $A$ be a data type. The following conditions on $A$ must be satisfied to
qualify for implementation in a relational database:
\begin{enumerate}
\item{$Con(A)$ must be defined. This means that the literal form of the
  data type must be provided or be trivial. For instance, the unit constants
  from \pq have a BNF style grammar definition, which qualifies as an
  intensional definition of $Con(\pqns)$.}
\item{$Dom(A)$ should be well defined, which means that is should be clear what
  its elements are. There should be a one-to-one correspondence between an
  object and its compositional properties. In other words, an object is
  completely defined by its compositional properties.}
\item{The \emph{compositional properties} of $A$ must be binary representable
  in the C language. This seemly arbitrary condition makes explicit that a
  translation from the recursive definitions of types, that are convenient and
  ubiquitous in the HL7 data types specification, must be translated to an
  implementable form.}
\item{The \emph{relational properties} of $A$ must be implementable as a
  User-Defined Function in the relational database.}
\item{Meaning of data values may not be lost or altered when the context in
  which it is used is changed. Specifically, its semantics may not change after
  the application of any number of relational operators.}
\end{enumerate}

\subsection{Nullflavors}
\label{ananull}
Since the nullflavors are extensions to each data type, the conditions of
section \ref{conditions} also apply here.

The Nullflavors extend the literal from of the underlying data type, i.e. for an
arbitrary data type $A$, $Con(A)$ is extended with the nullflavor constant
symbols, which are defined by extension.

\begin{figure}
\centering
\includegraphics[width=8.5cm]{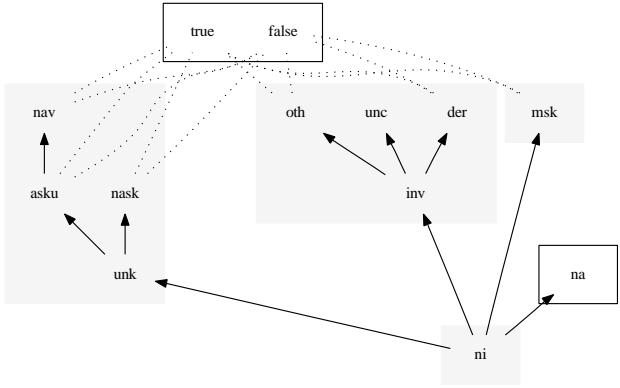}
\caption{Degree of knowledge ordering with nullflavors in boolean domain}
\label{fig:nfordering}
\end{figure}

Regarding conditions on $Dom(A)$, it is clear that Nullflavors serve the same
purpose of NULLs known from database literature, and it is likely that HL7s
nullflavor definitions originate from database literature. For instance, the
NULL in SQL:2003 has as meaning \emph{value at present unknown}, which
coincides with the nullflavor {\tt unk}. Also, the most general nullflavor {\tt
  ni} (no information) has the same name and meaning as the most general null
described by \cite{bjo91}. We believe the methodology developed in \cite{bjo91}
can be applied to the the nullflavors as well, for each data type.  The
nullflavors extend the literal form of the underlying data type, so the
nullflavor symbols are added to the set of constants {\bf con}. The most
general nullflavor {\tt ni} can be any object, no information is known. In the
case of all 9 nullflavors allowed on \blns, except {\tt ni} and {\tt na}, the
knowledge can be improved to the actual value. For instance, the knowledge of
{\tt nav} can be improved to be either true or false. Hence \evaluation{\tt
  nav} = $\{\rm true,\ false\}$. The nullflavor {\tt na} can never be improved,
  it is already perfect knowledge. Like \cite{zani} we extend the underlying
  domain $Dom(\blns)$ with the object described by \evaluation{\tt
    na}. I.e. $\rm Dom(\blns)=\{\rm true, false, \tt na\}$. The nullflavor {\tt
    ni} gives no information at all, hence \evaluation{\tt ni} = $Dom(\blns)$.
  The degree of knowledge expressed by nullflavors defines an ordering on the
  descriptions of the elements of $Con(\blns)$. The description \evaluation{\tt
    unk} = $\{\rm true, false\}$ is a better description than \evaluation{\tt
      ni}, since \evaluation{\tt unk} $\subset$ \evaluation{\tt ni}. True,
    false and {\tt na} are perfect descriptions, since the knowledge they
    describe cannot be improved. This is shown in figure \ref{fig:nfordering}.

A result of this analysis is that {\tt na} can never be improved to be either
\emph{true} or \emph{false}. This has consequences for the 11-valued BL logic
in cases where {\tt na} is involved. For instance, according to HL7, the
conjunction of \emph{true} and {\tt na} is {\tt na}. But it might as well be
\emph{false}, since {\tt na} cannot be improved to \emph{true}. The same holds
for disjunction. As an example, consider the following example with possible
situations for the propositions `Mary is pregnant' ($m$) and `Bob is pregnant'
($b$). Tables \ref{hl7pregnancy} and \ref{alterpregnancy} show the behaviour of
boolean operators involving {\tt na} as defined by HL7 and an alternative
behaviour, with the differences typeset italic. The tables also show
differences for logical combinations of {\tt na} and other nullflavors. The
rationale behind the differences for conjunction is that since a conjunction is
true iff both conjuncts are true, the conjunction is always false, since true
is not an element of \evaluation{\tt na}. For the disjunction of a nullflavor and
{\tt na}, the question is what answer gives the most information we know about
the result. Since the result can only `be made true' by the disjunct that is
not {\tt na}, if it is true, it is because of the not-{\tt na}
disjunct. Hence it makes sense to propagate the information this disjunct
carries in its nullflavor to the result of the disjunction. For instance,
suppose that the value of the proposition `Mary is pregnant' is masked,
nullflavor {\tt msk}. Under the alternative interpretation of $\lor$, the
disjunction `Mary is pregnant' or `Bob is pregnant' is {\tt msk}.

\begin{table}
\centering
\caption{HL7 {\tt na} behaviour}\vskip.5em
\begin{tabular}{|l|l|l|l|} \hline
$m$ & $b$ & $m \land b$ & $m \lor b$ \\ \hline\hline
true & na & na & true \\ \hline
false & na & false & na \\ \hline
asku & na & ni & ni \\ \hline
inv & na & ni & ni \\ \hline
msk & na & ni & ni \\ \hline
na & na & na & na \\ \hline
nask & na & ni & ni \\ \hline
nav & na & ni & ni \\ \hline
ni & na & ni & ni \\ \hline
oth & na & ni & ni \\ \hline
unk & na & ni & ni \\ \hline
\end{tabular}
\label{hl7pregnancy}
\end{table}

\begin{table}
\centering
\caption{Altered {\tt na} behaviour}\vskip.5em
\begin{tabular}{|l|l|l|l|} \hline
$m$ & $b$ & $m \land b$ & $m \lor b$ \\ \hline\hline
true & na & \emph{false} & true \\ \hline
false & na & false & \emph{false} \\ \hline
asku & na & \emph{false} & \emph{asku} \\ \hline
inv & na & \emph{false} & \emph{inv} \\ \hline
msk & na & \emph{false} & \emph{msk} \\ \hline
na & na & \emph{false} & \emph{false} \\ \hline
nask & na & \emph{false} & \emph{nask} \\ \hline
nav & na & \emph{false} & \emph{nav} \\ \hline
ni & na & \emph{false} & \emph{ni} \\ \hline
oth & na & \emph{false} & \emph{oth} \\ \hline
unk & na & \emph{false} & \emph{unk} \\ \hline
\end{tabular}
\label{alterpregnancy}
\end{table}

Regarding the binary representation of the nullflavors \emph{compositional
  properties}, it is easy to enumerate the 15 kinds into e.g. an integer or
nibble. The question however is in which structure in the database server to
store this integer. The extensible type system of the PostgreSQL relational
database was not meant to support more kinds of NULLs. Since it is open source,
we considered a number of alternatives, for instance a complete replacement of
the boolean {\tt isnull} in source code and the isnull bitmap in heaptuple
structure in memory and on disk by a set of nibbles. We did not choose this
option mainly because it would break compliance with SQL:2003 on standard NULL
handling. That meant that any implementation would be a hybrid solution of
database NULLs and HL7 NullFlavors. The question was to what extent there
should be an interaction between database NULLs and HL7 NullFlavors. For
instance, should {\tt NOT NULL} constraints on table columns mean that the data
values may not contain a nullflavor? And if a record contains a database NULL
value, should the HL7 {\tt isnull} relational property return true? We chose to
implement no interaction between database NULLs and NullFlavors at all. As a
consequence, any kind of nonnull constraints should be provided by the HL7
data type implementation itself. In the case of the \bn data type, HL7 shows it
is able to provide such constraints. In the case of the \ii data type however,
we believe that a nonnull flavor would be appropriate for use in a relational
database. See section \ref{instanceidentifier} for a discussion. The nibble
representing the nullflavor or 0 for nonnull, is added to the binary
representation of every implemented data type.

The \emph{relational properties} like {\tt isnull}, {\tt nonnull} are trivial
to implement with user defined functions.

The last condition that meaning of data values do not change under the
application of relational operators, poses no problems for most nullflavors,
but there might be a problem with the {\tt oth} nullflavor. If a data value has
the nullflavor {\tt oth} this means that the actual value is not a member of
the set of permitted data values in the constrained value domain of a
variable. This may occur when the value exceeds some constraints that are
defined in too restrictive a manner. For example, if the value for age is a 100
years, but the constraining model specifies that the age must be less than 100
years, the age may still be given, provided that the model does not make the
attribute mandatory.

One problem is the literal form of this value is not defined, though we could
choose, for instance, to write {\tt 100 yr NullFlavor.OTH} \footnote{The
  literal {\tt yr} was chosen as example. Actual UCUM units for year are {\tt
    a\_t}, {\tt a\_j} and {\tt a\_g}.}. Second, what is the status of this
object in $Dom(\pqns)$? By the definition of the \emph{equal} property, it is
equal to {\tt 100 yr}. But this is in conflict with the unique names
assumption. Also, depending on the way the constraint is specified, the meaning
of the nullflavor is altered under projection. If the constraint the data value
does not satisfy is defined on a relation attribute with a check constraint
like {\tt CHECK (age < '100 yr'::pq)}, it is known that {\tt oth} means that
the value violates this constraint. But projection of the relation attribute
`loses' the check constraint, and hence the meaning of the data value is
altered. We solved these issues by having data values with the other nullflavor
not compare with any other data value, so they do not violate any check
constraint except being nonnull or not the other nullflavor.

\subsection{Type hierarchy}
The internal representation of each subtype is binary compatible with its
direct super type on all components defined by the super type. Let $A, B$ be
data types. If $A$ and $B$ are different data types, a \emph{cast} is a function
of type $A \to B$. For each two data types $A, B$ where $A$ is a specialization
of $B$, we implemented two casts $A \to B$ and $B \to A$ that allow the data
values to be converted.

This allows the creation of a table with an attribute of type \any, that can
then hold data values of proper specializations of \any such as \pq.

\begin{verbatim}
CREATE TABLE testany (a hl7.any);
INSERT INTO testany VALUES ('10 ml'::pq);
SELECT * FROM testany;
   a
-------
 10 ml
(1 row)
\end{verbatim}

\subsection{Booleans}
Both boolean data types \bl and \bn satisfy all conditions of section
\ref{conditions} and the implementation was trivial aside from implementing the
nullflavors. One important aspect is that besides the casts from \bl and \bn to
\any and vice versa for the type hierarchy, we also implemented casts from and
to PostgreSQL's native data type {\tt boolean}. The bulk of the relational
properties on HL7 data types have either \bl or \bn as domain. The casts to {\tt
  boolean} enable that user defined functions implementing the relational
properties, may be used in filter expressions.

\subsection{Real Number}
\label{anareal}
A BNF-style grammar is defined by the standard for the literal form of reals,
so the first condition on $Con(real)$ is met. $Dom(real)$ is also well
defined, since it is $\mathbb{R}$.

Current CPUs provide IEEE-754 floating point arithmetic, making for an obvious
binary representation. The double precision numbers
have a precision of more than 15 fractional digits, which seems adequate for
the use cases with the HL7 real number data type.

However, as the following example calculation shows, IEEE-754 is not adequate
for the HL7's real. Comparing {\pqns}s that have comparable but different units
means computing and comparing canonical values which may result in rounding
errors\footnote{The rounding error $\epsilon$ is inherent in the IEEE-754
  specification. The actual value of $\epsilon$ differs per
  implementation. $\epsilon$ on Intel x86 hardware is $\approx$ $1.08e-19$.}  .

\begin{verbatim}
/* canonical values of 1l and dm3 differ */
SELECT 0.1^3::float8 - 0.001::float8 AS error,
       0.001::float8 = 0.1^3::float8 AS equal;
        error         | equal 
----------------------+-------
 2.16840434497101e-19 | f
\end{verbatim}

The rounding errors described above can be avoided with an arbitrary precision
real implementation.

\subsection{Physical Quantity}
\label{anapq}
The literal form $Con(\pqns)$ is well defined; it is the string representation of
the value followed by optional white space, followed by a unit string that
conforms to UCUM\cite{ucum}, which defines a BNF-style grammar for the unit
constants.

The domain $Dom(\pqns)$ poses no problems either, as it consists the union of all
data values of each unit, where the set of data values of a single unit is
isomorphic to the real numbers.

The binary representation of compositional properties are trivial, except the
value property, which is specified to be a real number. Section \ref{anareal}
discusses binary representations of real numbers.

A philosophical question arises when considering the following: is {\tt 1 m}
equal to {\tt 100 cm}? To answer this, it is necessary to think of what object
is described by the term {\tt 1 m}. Is this the same object as {\tt 100 cm} and
are both terms just different names for the same object? Or do both terms
describe two distinct objects? It is a choice that can be made, and there is no
wrong or right choice. Both the HL7 specification and conventions in database
theory make the choice that the terms describe different objects. From database
theory it follows from the unique names assumption. The HL7 specification does
so by making the distinction between the {\tt equal} and {\tt identical}
relational properties on \pqns. Two \pq objects are identical iff all
compositional properties equal. The {\tt equal} relation is not the usual
equality on objects, but an equivalence relation {\tt identical} relation on
the powerdomain of \pqns, which is defined as `has the same canonical value'.

\begin{verbatim}
SELECT equal('1m'::pq,'100cm'::pq);
 equal 
-------
 true

SELECT identical('1m'::pq,'100cm'::pq);
 identical 
-----------
 false
\end{verbatim}

The distinction between equality and identity has consequences for the index
access methods on the \pq data type. According to the HL7 specification, the
{\tt identical} property has no use other than in the definition of discrete
sets. In relational database terms, discrete sets can be constructed using
unique indexes. With index operator classes defined as follows, the user may
choose whether or not the unique names assumption holds.

\begin{verbatim}
CREATE OPERATOR CLASS pq_ops_equal
  DEFAULT FOR TYPE pq USING btree AS
    OPERATOR  3  = ,
    FUNCTION  1  btpqcmp_equal(pq, pq);

CREATE OPERATOR CLASS pq_ops_identical
  FOR TYPE pq USING btree AS
    OPERATOR  3  == ,
    FUNCTION  1  btpqcmp_identical(pq, pq);

/* default class uses equality */
CREATE INDEX idx ON rel
  USING btree(a);

/* choose identical operator */
CREATE UNIQUE INDEX idx ON rel
  USING btree(a pq_ops_identical);
\end{verbatim}

The following examples of the \pq data type show that it is easy to perform
calculations on physical quantities. Combined with the expression syntax of SQL
this results in concise source code that is easy to write and understand.

\begin{verbatim}
CREATE TABLE obs
  (ptnt int, effectivetime ts, dosage pq);
INSERT INTO obs VALUES
  (1, '200910011214', '10 ml'),
  (1, '200910041307', '100 ml'),
  (2, '200910080856', '1000 ml'),
  (1, '200910010915', '10 ml'),
  (3, '200910022312', '50 ml'),
...

SELECT ptnt,
       effectivetime::date,
       convert(SUM(dosage),'l') AS "sum",
       convert(AVG(dosage),'l') AS "average"
FROM obs
GROUP BY ptnt,effectivetime::date
HAVING contains('[100ml;500ml[', sum(dosage))
ORDER BY ptnt,effectivetime;
\end{verbatim}
\begin{verbatim}
 ptnt | effectivetime |  sum   | average 
------+---------------+--------+---------
    1 | 2009-10-01    | 0.12 l | 0.01 l
    1 | 2009-10-02    | 0.15 l | 0.05 l
    2 | 2009-10-02    | 0.3 l  | 0.05 l
    3 | 2009-10-01    | 0.15 l | 0.01 l
    3 | 2009-10-02    | 0.45 l | 0.05 l
    4 | 2009-10-04    | 0.3 l  | 0.1 l
(6 rows)
\end{verbatim}

\subsection{Flavors of physical quantities}
HL7s \emph{flavors} are data types that are defined by constraining another
data type. For instance, a \cou{pq\_time} is a physical quantity where the unit
compares to seconds. The {\tt CREATE DOMAIN} command of PostgreSQL provides
exactly this feature.

\begin{verbatim}
CREATE DOMAIN pq_time AS pq
        CONSTRAINT pq_time_compares_to_s
        CHECK (compares(value,'s'));

SELECT '10 ml'::pq_time;
ERROR:  value for domain pq_time violates
        check constraint "pq_time_compares_to_s"
\end{verbatim}

\subsection{Instance Identifier}
\label{instanceidentifier}
Like in the boolean case, it is easy to see that the instance identifier
data types satisfies the conditions from section \ref{conditions}.

An important aspect of relational databases is entity integrity, which states
that no primary key value of a base relation, which are those relations defined
independently of other relations, is allowed to be null or to have a null
component. Thus, if \ii is used to identify entities, it may not have a null
value. Unlike \blns, \ii has no flavor that constrains it to be nonnull. It is
neither possible to enforce this constraint on HL7 nullflavors with the
database NOT NULL constraint, since we implemented no interaction between the
database NULL and HL7s nullflavor, described in section \ref{ananull}. In other
words, it is impossible to force \ii data values to not have a nullflavor. A
solution would be to introduce the data type or flavor \cou{in}, analogue to the
\bn flavor of \bl.

\subsection{Interval PQ}
Although the literal form is specified, parsing intervals of physical
quantities is complicated by two HL7 definitions; the dash form is allowed in
\ivlpq and UCUM units can also contain brackets, as is shown by the following
two examples respectively:

\begin{verbatim}
SELECT '-8m--2m'::ivl_pq;
  ivl_pq   
-----------
 [-8 m;-2 m]

SELECT '[100mm[Hg];120mm[Hg]]'::ivl_pq;
  ivl_pq
-----------
 [100 mm[Hg];120 mm[Hg]]
\end{verbatim}

These forms do make parsing of literal forms more complicated; HL7
specification writers can make the implementers job easier if they separate
different items in a literal form by characters not allowed in those
items.

What are the elements of $Dom(\ivlpqns)$? Is \evaluation{{\tt <1m}} the same
object as \evaluation{{\tt ]NullFlavor.NINF m;1m[}}? The answer to this
question depends on the meaning of the \emph{same} predicate. The {\tt equal}
property is implemented as user defined function that is true if the low and
high values of both intervals are the same, and {\tt identical} returns true
iff they are equal and have the same literal form.

\begin{verbatim}
SELECT equal('30m [20m]'::ivl_pq,
             '[20m; 40m]'::ivl_pq);
 equal 
-------
 true
(1 row)
\end{verbatim}

\subsection{TS and interval of TS}
Like \ivlpqns, intervals of timestamps inherit attributes from the
\texttt{ivl$<$t$>$} template and thus have a clear literal definition.

The binary representation of time can be implemented as an offset to a common
epoch. Intervals can be seen as a two of these offsets when the bounds are
known, and a width if not. All these pose no problem when implementing the
compositional properties as a user defined type.

The HL7 specification defines a number of relational properties on \ts and
\ivltsns. Promotion and demotions are relations between quantity data types and
their interval data types. Since \ts have a certain precision, it can be viewed
as an interval of time. This interval is called the promotion of the \ts. Take
for instance the promotion of the \ts {\tt 2008} to the entire year 2008:

\begin{verbatim}
SELECT '2008'::ts, promotion('2008'::ts),
       demotion(promotion('2008'::ts));
  ts  |  promotion  | demotion
------+-------------+----------
 2008 | [2008;2009[ | 2008
(1 row)
\end{verbatim}

Other useful functions on \ivlts are the containment and overlap functions,
that are often used in filter expressions. Here is an example with the {\tt @>}
containment operator, that queries all records with a time interval that
contains the time interval {\tt 2001..2002}.

\begin{verbatim}
SELECT v FROM rel WHERE v @> '2001..2002';
      v
-------------
 [2000;2003[
 [2000;2004[
(2 rows)
\end{verbatim}

\subsection{Concept Descriptor}
The most used data type in the HL7 specification is the concept descriptor.  HL7
defines three forms of concept descriptors: \cdns, \cv and \csns. The standard
defines the literal form only for \csns. We adopt a common used literal form
like {\tt EVN:2.16.840.1.113883.5.1001}, for the event code in the {\tt
  ActMood} codesystem, and added valuesetoid, versions of codesystem and
valueset, and the originaltext.

Since user input of concepts using the standard literal form is error prone
because of the OID numbers involved, we also allow input of concepts using an
alternative format. To this end, we implemented a type modifier on the \cv
data type, that, if used, denotes the \emph{conceptdomainname}. For instance:

\begin{verbatim}
WITH cvvaluetable AS
(SELECT 'active|Ongoing treatment'::cv('ActStatus')
 AS c)
SELECT code(c), codesystem(c), codesystemname(c),
       codesystemversion(c), valueset(c),
       valuesetname(c), valuesetversion(c),
       originaltext(c)
FROM   cvvaluetable;
-[ RECORD 1 ]-----+-----------------------------
code              | active
codesystem        | 2.16.840.1.113883.5.14
codesystemname    | ActStatus
codesystemversion | 2009-08-30
valueset          | 2.16.840.1.113883.1.11.15933
valuesetname      | ActStatus
valuesetversion   | 2009-08-30
originaltext      | Ongoing treatment

\end{verbatim}

The type modifier format of the \cv data type may also be used to create
table attributes, which then acts as a constraint on the attribute:

\begin{verbatim}
CREATE TABLE rel (code cv('ActStatus'));
INSERT  INTO rel VALUES ('x');
ERROR:  invalid code 'x' for codeSystem ActStatus
INSERT INTO rel VALUES ('completed');
SELECT * FROM rel;
                                        code     ..
-------------------------------------------------..
 completed:2.16.840.1.113883.5.14@2009-08-30:2.16..
\end{verbatim}

Regarding the domain, there are similar issues as with the domain definitions
of \pq and \ivltsns. What are the distinct objects in the concept descriptor
domain? It makes sense to define the elements to be the objects that are
isomorphic to \cvns's compositional properties, but under this interpretation there
is a problem with the translations of the concept into other codesystems. We
believe translations are not among the basic intrinsic compositional properties
of the concept descriptor. Translations are added to concept descriptions in
communications as a service from a sending communication service to the
receiver, but adding a translation can never mean that the actual described
concept, the underlying object, is a different one.

An important property of the concept descriptor is the {\tt implies} property
with operator {\tt {<}{<}}, which is defined as follows: $a$ {\tt {<}{<}} $b$
iff $a$ implies $b$, which means that $a$ is a specific kind of $b$. With the
\cv data type it is possible to query for all specializations of e.g. a
SNOMED-CT clinical finding code {\tt 404684003}:

\begin{verbatim}
SELECT displayname(conceptid) from rel
WHERE conceptid <<
   '404684003|Clinical finding'::cv('SNOMED-CT');
                displayname                
-------------------------------------------
 Rupture of papillary muscle               
 Unspecified visual field defect           
 Oesophageal body web                      
 Benign tumour of choroid plexus           
 Sulphaguanidine adverse reaction          
..
\end{verbatim}

The last condition of section \ref{conditions}, which states that the meaning of
data values must be closed under application of relational operators, is not
satisfied by the \cs data type as it is specified. The conceptdomain is `fixed'
by the context in which the data value occurs. It has a literal form that
consists only of the code; other properties are omitted. In a relational
database setting, contexts of data values change under the application of
relational operators, and so will a possible fixed codesystem, if it exists.

Though \cs as it is defined does not comply with closure under relational
operators, the way it is used in the HL7 specification much resembles our
implementation of the \cv data type in its type modifier form, such as {\tt
  cv('ActMood')}. In the literal form input, only the code may be listed, and
when used in table attributes, the type modifier `fixes' the conceptdomain. But
in the \cv case, if e.g. the data value is projected into another tuple, no
context information is lost, since that is carried by the modified \cv type.

This concludes the analysis of the data types.

\pagebreak
\begin{figure}[p!]
\includegraphics[width=8.5cm]{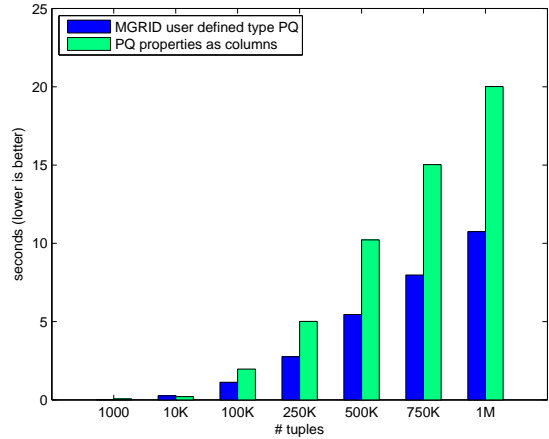}
\caption{Insert time}
\label{fig:inserttime}
\end{figure}

\begin{figure}[p!]
\includegraphics[width=8.5cm]{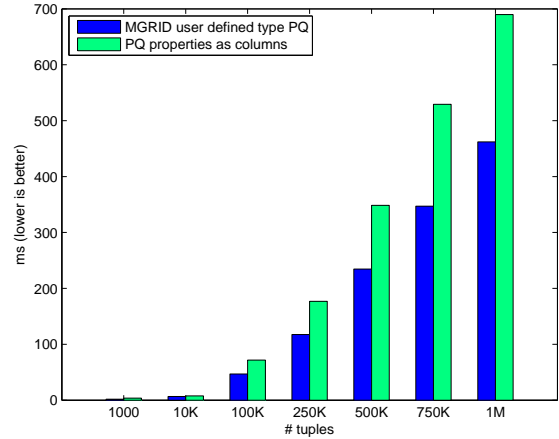}
\caption{Sequential scan}
\label{fig:seqscantime}
\end{figure}

\begin{figure}[p!]
\includegraphics[width=8.5cm]{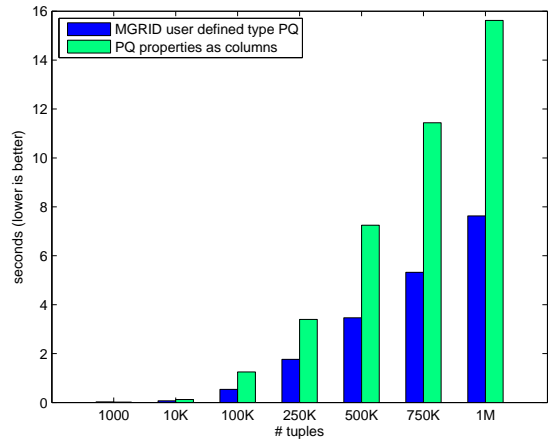}
\caption{Index creation time}
\label{fig:indexcreatetime}
\end{figure}

\begin{figure}[p!]
\includegraphics[width=8.5cm]{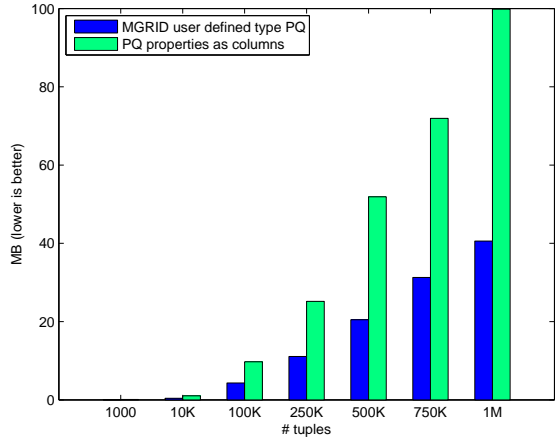}
\caption{Index size}
\label{fig:indexsize}
\end{figure}

\begin{figure}[p!]
\includegraphics[width=8.5cm]{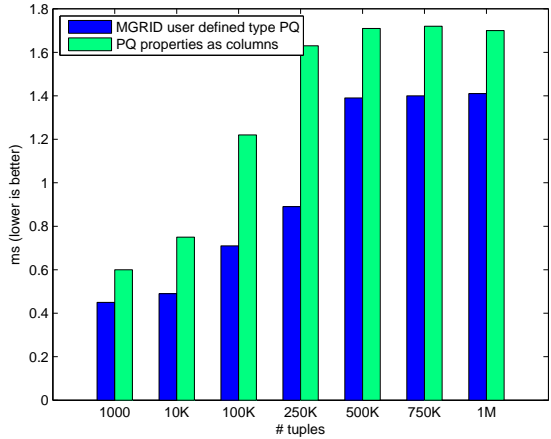}
\caption{Index equality scan}
\label{fig:indexscan}
\end{figure}

\begin{figure}[p!]
\includegraphics[width=8.5cm]{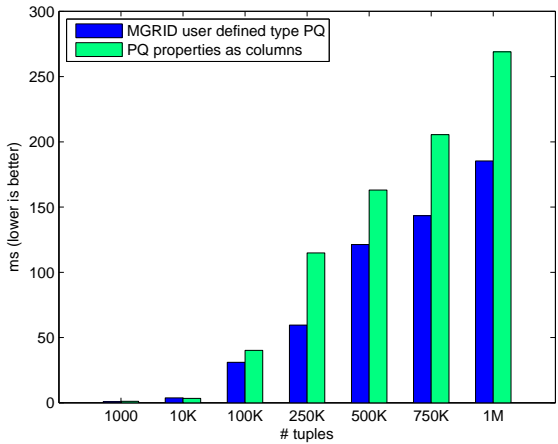}
\caption{Index range scan}
\label{fig:indexrangescan}
\end{figure}

\section{Performance}
This section shows the results of a number of tests that were performed on
synthetic data, were two implementations of the physical quantity data type are
compared:
\begin{enumerate}
\item{\emph{MGRID user defined type PQ} uses the user
defined \pq data type as described in this paper.}
\item{\emph{PQ properties as columns} uses the standard types of the database
  server, where the compositional properties are distinct columns in the test
  relation.}
\end{enumerate}

The graphs show the average of 20 test runs for each different number of tuples.
During each run, new random data was generated with the following
characteristics:

\begin{itemize}
\item{20 different units chosen at random and always the unit {\tt m}. We
  believe 20 different units is a reasonable ballpark figure for common uses of
relations with \pq values. The unit {\tt m} was chosen to match queries in the
performance test.}
\item{for each unit, values were randomly distributed in a Gaussian
  distribution with $\mu$ 0 and $\sigma$ 10000. The size of the distribution
  was chosen to match the range queries to match a small fraction of the total
  number of tuples. The Gaussian distribution was chosen because it matches
  many natural occurring phenomena.}
\end{itemize}

Figure \ref{fig:inserttime} shows the amount of time it took to create the
test relations. In both cases the same UCUM parser validated correctness of unit
strings and comparable calculations were performed to insert
proper values into the database.
Figure \ref{fig:seqscantime} shows a sequential scan that queries for values
between {\tt 1.0km} and {\tt 1.2km}.
The user defined type \pq is between 30\% to 50\% faster in these operations,
which is probably due to the SQL function call and context switch overhead that
is expected in the implementation with properties as columns, as well as the
larger number of columns involved in the latter case.

Figures \ref{fig:indexcreatetime}, \ref{fig:indexsize}, \ref{fig:indexscan} and
\ref{fig:indexrangescan} compare index performance in creation time, size and
equality and range scanning. The equality scan selected the value {\tt 1.2km}
and returned at least a single row. The range scan selected all values between
{\tt 1.0 km} and {\tt 1.2km} and returned about 0.1\% of all tuples.  The
graphs show that the user defined type outperforms the implementation based on
a set of distinct table columns in both size of the index as well as index
scans. The equality scan does not exhibit a clear difference between some of
the data sizes. We believe this is due to the fact that these timings differ
tens of microseconds, and are below the point where our timing method can
produce accurate results. It is expected that if a more accurate measurement
for short timing could be used, these values would be in tune with the timings
for the larger relation sizes. The timings for the index range scan may seem
large relative to the equality scan. It must be kept in mind however that the
scan must skip values that are not of the same unit and that there is nullflavor
handling taking place. For instance, a value of {\tt Nullflavor.TRC ml} must be
returned in a query that requests values {\tt > 0 ml}. These properties make
the index structure more complex.

On the whole these figures show that the user defined type \pq outperforms an
implementation based on a set of distinct columns that match \pqns's
compositional properties. With regard to index size, the user defined \pq type
is only 40\% of the size of the compositional \pq implementation. The user
defined type is about 40\% faster at index creation than the compositional
implementation. The consequences for insertion of tuples under the presence of
indexes are similar. With index scanning the performance gain of using the user
defined \pq is about 25\% for range queries and comparable for equality
queries.

\section{Conclusions}
Though the HL7 data types are not designed to be implemented in a relational
database, it is possible to implement several of the most used data types such
as \pqns, \blns, \ii and \cvns.

The type hierarchy posed no problem to implement. Other aspects of the data
type specification had a less obvious solution, such as nullflavor integration,
defining missing literal forms, and choice in application of the unique names
assumption. Some data type constructs could be implemented in a single database
data type, but are more appropriately mapped to multiple data types or even
relations instead of single data values. Good examples here are the
translations of concepts in the concept descriptor data type, and discrete
sets. The \emph{other} nullflavor and the \cs data type are problematic in the
sense that their meaning depends on the context in which a data value is
used. None of these problems were show stoppers.

The relational properties are easy to implement with user defined
functions. Expressions involving HL7 predicates may be used in filter
expressions, after adding casts that allow interaction between the database
boolean and HL7's \bl and \bnns. It is possible to provide indexing for the
complete range of relational operators on data types.

Though the HL7 specification defines names for relational properties, it lacks
operator symbols such as {\tt +} for addition. As a result, differences in
operator symbol designations can be expected between implementations of the HL7
data types.

Our analysis results in the following suggestions for improvement in the HL7
data type specification:
\begin{itemize}
\item{Add a literal form for all data types that do not yet have one defined.}
\item{Translations of a concept descriptor should not be part of the
  concept descriptor data type.}
\item{Remove constructs where the meaning of an artifact depends on the context
  in which it is used.}
\item{An alternative interpretation of binary operations involving
  the \emph{not applicable} nullflavor is proposed in table
  \ref{alterpregnancy}.}
\item{Add a note that IEEE-754 floating point representations found in hardware
  co-processors is inadequate for calculations involving physical quantities.}
\item{Add a nonnull flavor of the \ii data type.}
\item{Add operator symbols for relational properties.}
\end{itemize}

It may seem old fashioned to implement HL7 data types as user defined types;
current practice is to implement complex types in the persistence layer, where
the database is used to store the individual components of these types, using
object relational mapping (ORM). We believe that augmenting the database server
by adding these types as native type makes sense for several reasons. First, it
allows the database server to store and retrieve information more effectively,
thereby reducing the amount of data exchanged with the persistence layer and
providing better scaling. Besides a smaller performance footprint of the
persistence layer, the user defined types also provide powerful primitive
operations on HL7 data types to the persistence layer, thereby reducing its size
and complexity and indirectly the prevalence of software engineering
errors. Third, consistent reasoning about HL7 data in a heterogeneous
application environment requires a consistent common understanding between the
applications, that the database server with user defined types is in a unique
position to supply.

Our approach does not rule out ORMs, but does simplify them by making use of
the database HL7 data types. We have shown that operations on the user defined
types are significantly faster and occupy less space than an alternative
implementation that resembles current practice of implementing complex types.

\bibliographystyle{abbrv}

\vfill

\appendix
\section{MGRID}
The mission statement of MGRID, a Dutch IT company, is to develop the
ultimate medical database. Part of the database solution of MGRID is a
structured HL7 RIM database. This database makes use of the HL7 data types that
are the subject of this paper. Other aspects of MGRID's database solution
encompass solutions for high availability, linear scalability, performance,
encryption of private medical data and authorization for users based on their
role as known in the RIM database. MGRID also provides fast terminological
services and tooling for model driven architectures with Detailed Clinical
Models (DCM).

The MGRID RIM database can be used in conjunction with object relational
mapping (ORM) tools like hibernate. In this combination MGRID guarantees
performance and delivers powerful primitive operations on medical data in the
HL7 and SNOMED-CT languages. The mapping with hibernate ensures that the
application software engineer can make optimal use of MGRID's primitives in his
or her own language.
\end{document}